\documentclass[twocolumn,prl,showpacs,preprintnumbers,amsmath]{revtex4}
\usepackage{graphicx}
\usepackage{dcolumn}
\usepackage{bm}

\begin{document}

\title{Fermi surface shrinking and interband coupling in iron-based pnictides}

\author{L. Ortenzi$^{1,2}$}

\author{E. Cappelluti$^{2,1}$}

\email{emmanuele.cappelluti@roma1.infn.it}

\author{L. Benfatto$^{3,1,2}$}

\author{L. Pietronero$^{1,2}$}

\affiliation{$^1$Dipart. di Fisica, Universit\`a ``La Sapienza'',
P.le A. Moro 2, 00185 Rome, Italy}

\affiliation{$^2$SMC Research Center, CNR-INFM, c/o ISC-CNR, via dei
Taurini 19, 00185 Rome, Italy}

\affiliation{$^3$Centro Studi e Ricerche ``Enrico Fermi'',
v. Panisperna 89/A, 00184,
Rome, Italy}

\begin{abstract}
Recent measurements of Fermi surface with de Haas-van Alphen oscillations
in LaFePO showed a shrinking of the Fermi pockets with respect to
first-principle calculations, suggesting an energy shift of the hole
and electrons bands with respect to local density approximation.
We show that these shifts are a
natural consequence of the strong particle-hole asymmetry of electronic
bands in pnictides, and that they provide an indirect experimental evidence
of a dominant interband scattering in these systems.
\end{abstract}

\maketitle

A new challenge in the field of condensed matter is represented by the
recent discovery of high-$T_c$ superconductivity in the iron-based pnictide
family \cite{kamihara}.  First-principle calculations have soon identified
four main bands at the Fermi level with two-dimensional character: two
hole-like pockets around the $\Gamma$ point and two electron-like pockets
around the M point \cite{lebegue,singh1,singh2,nekrasov,boeri,mazin}.  A
fifth three-dimensional band crossing the Fermi level at the $\Gamma$ point
is sometimes predicted by Local Density (LDA) calculations
\cite{lebegue,singh1}, even if its position with respect to the chemical
potential is strongly affected by the interlayer distance.  Since
determining the Fermi surface topology is the first step toward the
understanding of these new materials, a lot of experimental work has been
devoted to its investigation. A momentum-resolved mapping of the dispersion
of the occupied quasi-particle states in the normal state and in the
superconducting one has been provided by angle-resolved photoemission
spectroscopy (ARPES), both in 1111 \cite{lu} and 122 systems
\cite{ding1,ding2,yang,wray,yi}. An alternative technique probing the Fermi
surface topology, which is not momentum resolved but has the advantage of
being a bulk probe, is based on de Haas-van Alphen (dHvA) magnetization
measurements, which allow one to estimate the size of the Fermi areas and
the effective mass $m^*$ for each Fermi sheet
\cite{coldea,sugawara,carrington,sebastian,analytis}.  Although an overall
qualitative agreement is found between LDA calculations and these
experiments, interesting discrepancies still remain.  In general, we can
distinguish between {\em high energy} and {\em low energy}
discrepancies.  ARPES data in Ba$_{0.6}$K$_{0.4}$Fe$_2$AS$_2$
\cite{ding1,ding2} and LaFePO \cite{lu} for instance find overall
bandwidths a factor 2 smaller than the LDA calculations, possibly related
to the static electronic correlation \cite{sadovskii}.  On the other hands,
the ARPES observation of Fermi velocities a factor 4 larger than the LDA
calculated low-energy ones \cite{ding2} points out a relevant role of
dynamical quantum renormalization effects associated with a low-energy
boson-mediated interaction.

Besides the bandwidth renormalization, a second striking result emerging
from dHvA is a substantial shift of the bands with respect to the Fermi
level when compared to LDA calculations. Indeed, the accurate determination
of the Fermi surface areas provided by dHvA gives values smaller then those
expected by LDA. Such discrepancy can be accounted for assuming a shift of
the LDA bands.  Notably, such
shifts have different signs in hole and
electron bands, being
{\rm downward} for the hole-band and {\em
upward} for the electron ones\cite{coldea,carrington}. This effect
seems to persist to higher energies, as one can infer from ARPES, which
gives a systematical reduction of the energy difference between the bottom
of the electron bands and the top of the hole bands with respect to the
values predicted by LDA \cite{ding2,yang,wray,yi}. 
Even though a certain degree of inaccuracy could be present
in density-functional theory calculations, the persistent observations
of such shifts suggests a robust feature in these materials
whose origin is at the moment unknown.

In this Letter we present a comprehensive explanation for the origin of the
band shifts observed in dHvA experiments.  We show that they
are a natural consequence of the multiband character of these systems and
of the strong particle-hole asymmetry of the bands, which induce
finite-band self-energy effects that are usually irrelevant for the half-filled
single-band case.  We also show that a simple analysis of the dHvA data
provides evidence that the dominant interaction in these
systems is the interband one between hole and electronic states, as it has
been already argued on the basis of the nesting between hole and electron
Fermi sheets \cite{mazin,ma,raghu,chubukov,kuchinskii}.  More precisely, our 
calculations give an estimate of the interband coupling $V$ of the order $V
\approx 0.46$ eV, which gives rise also to a mass enhancement
$Z_\alpha \approx 1.4$ for each band.

We start our analysis by discussing first
a generic multiband system, where the electrons are
coupled to a bosonic mode with local propagator $D(\omega_l)=
\int d\Omega 2\Omega B(\Omega)/(\Omega^2+\omega_l^2)$,
and where $B(\Omega)$ is the density of states of the
bosonic excitations.
The 
self-energy in the Matsubara space
can be written as:
\begin{eqnarray}
\Sigma_\alpha(i\omega_n)
&=&
-T\sum_{m,\beta}
V_{\alpha,\beta}
D(\omega_n-\omega_m)
G_\beta(i\omega_m),
\label{self}
\end{eqnarray}
where $\alpha, \beta$ are band indexes,
$V_{\alpha,\beta}$ ($V_{\alpha,\beta}=V_{\beta,\alpha}$)
is the multiband interaction, 
which for a retarded interaction is always positive
($V_{\alpha,\beta}>0$),
and $G_\alpha(z)$ is the local Green's function
for the $\alpha$ band, namely:
\begin{eqnarray}
G_\alpha(z)
&=&
\int_{E_{\mathrm {min},\alpha}}^{E_{\mathrm {max},\alpha}} 
d\epsilon N_\alpha(\epsilon)
\frac{1}{z-\epsilon-\Sigma_\alpha(z)+\mu}.
\label{green}
\end{eqnarray}
Here $N_\alpha(\epsilon)$ is the electronic density of states of the $\alpha$
band, with upper and lower band edges $E_{\rm max,\alpha}$ and $E_{\rm
min,\alpha}$, respectively, and we drop the spin index since it does not play
any role in the following.

In the conventional Eliashberg analysis \cite{eliashberg}, one usually
assumes that the distance of the chemical potential from the band edges is
much larger than the typical boson energy scale, so that one can
approximate the DOS with its value at the Fermi level,
$N_\alpha(\epsilon)\approx N_\alpha(\mu)$, and one can extend the
integration limits over $\epsilon$ in Eq.\ (\ref{green}) to $\pm \infty$.
In this way one is enforcing the particle-hole symmetry and the Matsubara
self-energy $\Sigma_\alpha(i\omega_n)$ is purely imaginary. 
Enforcing implicitly the particle-hole symmetry can be however quite
dangerous in systems as the iron-pnictides where each band is strongly away
from the half-filling.  As a byproduct of taking into account the
particle-hole {\em asymmetry}, the self-energy acquires a finite real part
$\chi_\alpha(i\omega_n) \equiv \mbox{Re}\Sigma_\alpha(i\omega_n)\neq 0$,
whose low energy limit $\chi_\alpha=\chi_\alpha(i\omega_{n=0})$ gives
rise to a band shift that can be in general different in each band. While
this effect is usually disregarded in single-band systems, because it can
be absorbed in a redefinition of the chemical potential, in the multiband
case it can lead to observable relative shifts of the various bands with
respect to the Fermi level.

Before undertaking a fully numerical solution of
Eqs. (\ref{self})-(\ref{green}), it is instructive to consider the result
of the lowest order perturbation theory,  where the Green's function
$G_\beta(i\omega_m)$ in Eq. (\ref{self}) is taken to be the non-interacting
one with $\Sigma_\beta(i\omega_m)=0$.  We shall
assume also purely two-dimensional parabolic bands, so that 
$N_\alpha=1/(E_{\rm max,\alpha}-E_{\rm min,\alpha})$ and we consider
for the moment, for simplicity, a Einstein mode
$B(\Omega)=(\omega_0/2)\delta(\Omega-\omega_0)$.
We get thus the analytical expression (for $T\approx 0$):
\begin{eqnarray}
\chi_\alpha
&=&
-\frac{\omega_0}{2}\sum_\beta V_{\alpha,\beta}N_\beta
\ln
\left|
\frac{\omega_0-\mu+E_{\rm max,\beta}}
{\omega_0+\mu-E_{\rm min,\beta}}
\right|.
\end{eqnarray}

In most cases, the exchanged boson energy is the lowest energy scale
in the system, so that
\begin{eqnarray}
\chi_\alpha
&\approx&
-\frac{\omega_0}{2}\sum_\beta V_{\alpha,\beta}N_\beta
\ln
\left|
\frac{E_{\rm max,\beta}-\mu}{E_{\rm min,\beta}-\mu}
\right|.
\label{chiapp}
\end{eqnarray}
Eq. (\ref{chiapp}) gives correctly $\chi_\alpha=0$ in the case of
particle-hole symmetry $|E_{\rm max,\beta}-\mu|= |E_{\rm
min,\beta}-\mu|$. On the other hand, in an electron-like band $|E_{\rm
max,\beta}-\mu|> |E_{\rm min,\beta}-\mu|$, while the opposite inequality
applies for a hole-like band.
Let us consider now a single band system.
In this case the band shift $\chi_{\alpha,0}$ is compelled
to be positive (upward) for a hole-system and negative
(downward) for an electron-like band.
Such observation seems at odds with what
reported in LaFePO.

Things are however more subtle in multiband systems,
where the shift of each
band depends also on the contribution of {\em all the other bands}
weighted with the corresponding interband coupling.
This effect is particularly important in pnictides
where the dominant coupling
is thought to be the spin-mediated interband interaction.
In such a situation, the particle-hole asymmetry of
the electron bands is responsible for the downward shift of
the hole-like bands which, vice versa, give rise to the
upward shift of the electron ones.
We would like to stress that these argumentations
do not depend on the band or interaction details.
In full generality {\em the simple
observation of a upward shift of the electron bands and
the downward shift of the hole-like ones is a direct
experimental evidence in these compounds of the dominance of the
interband coupling on the intraband one, which would produce
the opposite scenario}.

\begin{figure}[t]
\includegraphics[scale=0.16,clip=]{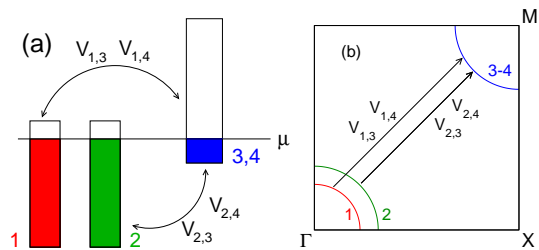}
\includegraphics[scale=0.18,clip=]{sketch4.eps}
\caption{(Color online). Sketch of our band structure model (a)
and of the corresponding Fermi surfaces (b). We also show the only non-zero 
interband couplings.}
\label{f-sketch}
\end{figure}

Within this framework, a more detailed analysis of the dHvA measurements in
LaFePO \cite{coldea} can provide an useful insight into the {\em strength}
of the interband coupling in pnictides.  Following Ref. \cite{lara} we
consider four bands ($1,2$ hole-like and $3,4$ electron-like) and we assume,
because of the nesting properties,
a purely interband scattering connecting hole and
electron bands, namely $V_{1,3}, V_{1,4}, V_{2,3}, V_{3,4}$, (see
Fig. \ref{f-sketch}).  Note that, unlike to other pnictides, all the Fermi
areas in LaFePO are quite comparable, so that we can expect that interband
scattering $V_{\alpha,\beta}$ does not depend much on the 
band index, and we can assume, as a first approximation,
$V_{1,3}=V_{1,4}=V_{2,3}=V_{3,4}=V$.  We model each band with a purely
two-dimensional parabolic dispersion, and we assume for simplicity bands
$3,4$ to be
completely degenerate. The band parameters are extracted from the LDA
calculations \cite{lebegue}. More explicitly, focusing on
the $k_z=0$ plane, we determine the Fermi vectors $k_{\rm F,\alpha}$ from
the $\Gamma$-X cut for the hole-bands, and from the M-X cut for the
electron ones.  Setting for convention the Fermi level $\mu=0$, we also
take from LDA calculations the nearest band edge for each band, namely
$E_{\rm max,1}$, $E_{\rm max,2}$ for the hole-like bands, $E_{\rm min,3}$,
$E_{\rm min,4}$ for the electron ones, and we use it to estimate the
non-interacting mass for each band $m_\alpha=\hbar^2 k_{\rm
F,\alpha}^2/2E_{\mathrm{max(min)},\alpha}$ and the corresponding density of
states $N_\alpha=m_\alpha a^2/2\pi \hbar^2$ ($a$ is here the in-plane
lattice constant). Finally, the effective band edge far from the Fermi
level \cite{note-edge} follows from the relation $N_\alpha=1/(E_{\rm
max,\alpha}-E_{\rm min,\alpha})$.  All the estimated values are reported in
Table \ref{t-table}.

\begin{figure}[t]
\includegraphics[scale=0.38,clip=]{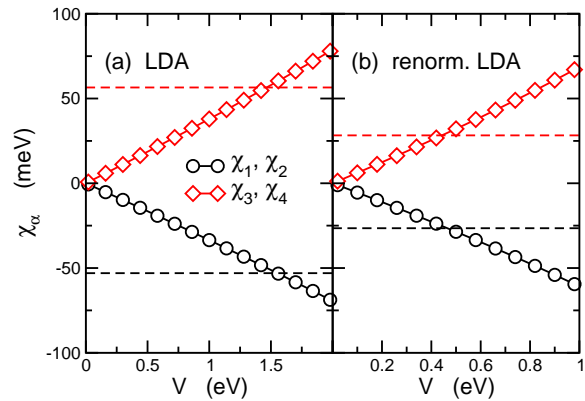}
\caption{(Color online). (a) Bands shift $\chi_{0,\alpha}$
as function of the interband coupling $V$ for
the parameter set taken from LDA calculations (table \ref{t-table}, top).
The horizontal dashed lines mark the average band shift
as estimated in Ref. \cite{coldea,carrington};
(b) same as in panel (a)
but considering LDA band structure renormalized by a factor 2 (table
\ref{t-table}, bottom).}
\label{f-plot}
\end{figure}

We can now employ a full numerical solution of
Eqs. (\ref{self})-(\ref{green}) to get a qualitative estimate the strength
of the interband coupling $V$ by comparing the calculated magnitude of the
band shifts and the effective mass $m_\alpha^*$
with the experiments.
This latter quantity can be evaluated
from the numerical solution of Eqs. (\ref{self})-(\ref{green})
as $m_\alpha^*=m_\alpha Z_\alpha$,
where $Z_\alpha=1-\mbox{Im}\Sigma_\alpha(i\omega_n)/\omega_n|_{n=0}$.
To account for a spin-mediated interaction mechanism we
use the Lorentzian spectrum typical of spin fluctuations\cite{millis}
$B(\Omega)\propto\Omega\omega_0/\pi(\omega_0^2+\Omega_2)$, with the
characteristic energy scale $\omega_0=20$
meV\cite{bang,matan,osborn,christianson}.
The smallness of $\omega_0$ compared with 
the Fermi energies, $E_{\rm F}\approx 100-200$ meV (Table \ref{t-table}),
guarantees the validity of Migdals' theorem \cite{migdal}
also for spin-fluctuations,
as discussed e.g. in Refs. \cite{millis,chubukov2,norman}.
\begin{table}[t]
\begin{center}
\begin{tabular}{ccccccc}
\hline \hline
band & 
$m_\alpha/m_e$ & $E_{\rm max,\alpha}$ & $E_{\rm min,\alpha}$
& $N_\alpha$ & $\chi_\alpha$ & $m_\alpha^*/m_e$  \\
index&  & (eV) & (eV) &  (eV$^{-1}$) & (meV) & 
\\
\hline
\multicolumn{7}{c}{LDA calculations LaFePO} \\
1   & 0.58 & 0.205 & -5.031 & 0.191 & -53 & 1.0 \\ 
2   & 1.14 & 0.205 & -2.462 & 0.375 & -53 & 2.0 \\
3,4 & 0.79 & 3.551 & -0.295 & 0.260 &  62 & 1.4 \\
\hline
\multicolumn{7}{c}{Renormalized LDA calculations LaFePO} \\
1   & 1.16 & 0.102 & -2.516 & 0.382 & -26.5 & 1.6 \\ 
2   & 2.28 & 0.102 & -1.231 & 0.750 & -26.5 & 3.2 \\
3,4 & 1.58 & 1.776 & -0.147 & 0.520 &  31   & 2.3 \\
\hline \hline
\end{tabular}
\end{center}
\caption{Microscopic band parameters extracted from unrenormalized LDA
calculations \cite{lebegue} (top), and renormalized ones (bottom). Also
shown are the calculated band shifts
$\chi_{0,\alpha}$ and the renormalized mass $m_\alpha^*$ for $V=1.55$ eV
and $V=0.46$ eV, respectively.  $m_e$ is the free electron mass.}
\label{t-table}
\end{table}
In Fig. \ref{f-plot}a we show the band shifts $\chi_{\alpha,0}$ evaluated from
the numerical solution of Eqs. (\ref{self})-(\ref{green})
as a function of $V$ by using
the set of microscopic parameters obtained by bare LDA calculations.  
From Fig. \ref{f-plot}a we get an estimate $V \approx
1.55$ eV to account for the hole band shift $\Delta_{1,2}\approx 53$ meV
reported in Ref. \cite{coldea}.
This value $V \approx 1.55$ eV gives also a theoretical positive
shift $\chi_3=\chi_4 \approx 60$ meV for the electron bands, which
agrees quite well with the average experimental value
$\bar{\Delta}_{3,4}=(\Delta_3+\Delta_4)/2=56.5$ meV\cite{coldea}.
It is clear that
possible differences between the set of hole and electron bands can arise
in a more refined treatment when the two electron bands are no more assumed
to be degenerate, and when interband coupling is not taken constant. This
value $V \approx 1.55$ eV would also give weak-coupling
multiband dimensionless interactions
$\lambda_{\alpha,\beta}=V_{\alpha,\beta}N_\beta$:
$\lambda_{13}=\lambda_{23}=\lambda_{14}=\lambda_{24}=0.40$,
$\lambda_{31}=\lambda_{41}=0.30$, $\lambda_{32}=\lambda_{42}=0.58$
which yield effective masses
$m_\alpha^*/m_e \approx 1, 1.4, 2$ smaller than
what reported in Refs. \cite{coldea,carrington,sugawara}.
A numerical solution of the multiband Eliashberg
theory would give also a critical temperature $T_c
\approx 29$ K, larger than the experimental one,
$T_c \approx 6$ K.

We identify the origin of this discrepancy in the lack, in the above
analysis, of high-energy band renormalization pointed out by ARPES
experiments\cite{lu}, which show that the real band structure is roughly
twice narrower than the LDA one.  The nature of such high-energy
renormalization is at the moment unknown, but it must be of different
origin than the interband spin-fluctuation coupling which acts on the
energy scale $\omega_0$. One can easily see indeed that for a retarded
interaction the band renormalization function
is negligible $Z(\omega)\sim 1$ when
$\omega\gg\omega_0$ \cite{eliashberg}.  This means that the responsible
mechanism for such high-energy renormalization must be operative on an
energy scale larger than the electronic bandwidth, suggesting once more a
correlation-driven origin \cite{sadovskii}.  Note also that in this case
such mechanism is not expected to affect the band shift since, as shown in
Eq. (\ref{chiapp}), $\chi_{\alpha,0}\approx 0$ for $\omega_0 \gg E_{\rm
max(min),\alpha}$.

\begin{figure}[t]
\includegraphics[scale=0.36,clip=]{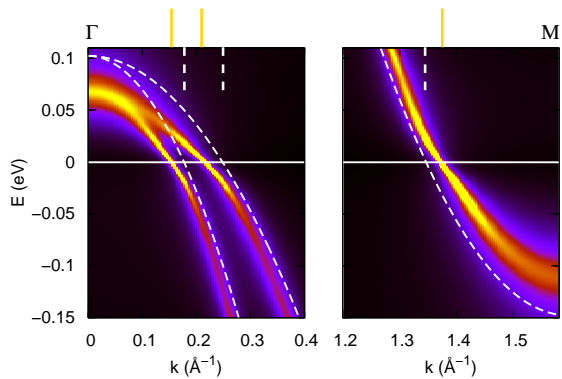}
\caption{(Color online). 
Intensity map of the spectral function for the
interacting hole bands (left panel) and the two degenerate
electron bands (right panel),
obtained by using LDA parameters renormalized
by a factor 2 and $V=0.46$ eV. The dashed lines
represent the non-interacting parabolic bands,
and the horizontal solid line is the chemical potential.
The ticks on the top mark the value of the 
Fermi vectors in the presence (solid yellow ticks) and in the 
absence (white dashed ticks) of interaction. Note the shrinking of the 
Fermi surfaces due to the coupling to the retarded bosonic mode.}
\label{f-map}
\end{figure}

We take into account the bandwidth narrowing by renormalizing the whole LDA
band structure as experimentally pointed out by ARPES in LaFePO and
Ba$_{0.6}$K$_{0.4}$Fe$_2$AS$_2$.  The corresponding microscopic parameters
in our model are listed in Table \ref{t-table}. Note that, in such rescaled
band structure, also the band shifts needed to recover the experimental
Fermi areas measured by dHvA techniques are reduced by a corresponding
factor 2. The plot of the band shifts as function of the interband
coupling, for such renormalized LDA band structure, is shown in
Fig. \ref{f-plot}b. Our estimate of the interband interaction in this case is
thus $V \approx 0.46$ eV, which gives weak-coupling values
$\lambda_{13}=\lambda_{23}=\lambda_{14}=\lambda_{24}=0.24$,
$\lambda_{31}=\lambda_{41}=0.18$, $\lambda_{32}=\lambda_{42}=0.34$,
with a $T_c \approx 9$ K and
$m_\alpha^*/m_e\approx 1.6-3.2$, in better agreement with
dHvA \cite{coldea,carrington,sugawara} and
specific heat measurements\cite{carrington,mcqueen,serafin}.
The corresponding intensity map of the
spectral function for the interacting hole and electron bands, as obtained
by the Marsiglio-Schossmann-Carbotte analytical continuation \cite{msc}, is
also shown in Fig. \ref{f-map} along with the bare band dispersions
renormalized by the factor 2.
Here it is clearly visible the shrinking of the Fermi area of each band due
to the interband self-energy effects ($\chi_\alpha$) which give rise to a
downward shift of the hole bands and to an upward shift of the electron
ones.  Note that such self-energy corrections due to the particle-hole
asymmetry survive until energies much larger than $\omega_0$, so that the
energy shift is effective also at momenta far from the Fermi level. In
particular this means that, when interband interactions are predominant,
the top of the hole bands (at $\Gamma$) and the bottom of the electron
bands (at $M$) approach each other with respect to the prediction of LDA,
in agreement with the ARPES observation in 122
compounds\cite{ding2,yang,wray,yi}.  Note also in Fig.\ \ref{f-map} the
effective mass renormalization at the Fermi velocity
$m_\alpha^* =m_\alpha Z_\alpha$ due to the retarded interband
interaction.  This effect, unlike the band shift $\chi_\alpha$, disappears
at the scale energy $\omega_0$, giving rise to a ``kink'' in the electronic
dispersion, recently observed in pnictides \cite{wray} and widely discussed
in the past in the context of cuprates \cite{damascelli}.

In conclusion, we demonstrated that the band shifts reported in pnictides
when comparing the experimentally measured Fermi surfaces and band
dispersions with LDA calculations are a direct consequence of the coupling
to a bosonic mode, once that the strong particle-hole asymmetry and the
multiband character of these systems are properly taken into
account. Moreover, we showed that the sign of the measured shifts, with
both the hole and electron bands approaching the Fermi level, suggests
that interband interactions dominate over interband
ones.  Our results pose strong constraints on the
theoretical modeling of interactions in pnictides, and show unambiguously
that finite-band effects cannot be disregarded in these systems.

This work was partially funded by MIUR
project PRIN 2007FW3MJX.


\begin{thebibliography}{99}
\bibitem{note-edge} 
It is clear that this value does not represent the true band edge as given
in LDA, but an effective one needed to guarantee the 
total number of states.
\end{thebibliography}
\end{document}